\documentclass[USenglish,twocolumn]{article}	

\usepackage[utf8]{inputenc}				
\usepackage[big,online]{dgruyter}	
\usepackage{lmodern} 
\usepackage{microtype}
\usepackage[numbers,square,sort&compress]{natbib}
\usepackage{graphicx}
\usepackage{dcolumn}
\usepackage{lipsum}
\usepackage{bm}
\usepackage{tabu}
\usepackage[utf8]{inputenc}
\usepackage[T1]{fontenc}
\usepackage{xcolor}
\usepackage{amssymb}
\usepackage{comment}
\usepackage{soul}

\newcommand{\RH}[1]{\textcolor{blue}{ #1}}
\newcommand{\dd}[3]{\ensuremath{\frac{d^\text{#3} #1}{d #2^\text{#3}}}}

\newcommand{\hatt}[1]{\ensuremath\hat{{\mathbf #1}}}

\newcommand{\AVW}[1]{\ensuremath{{\bf AVW}}}

\theoremstyle{dgthm}

\theoremstyle{dgdef}

\begin{document}

	\articletype{Research Article}
	\received{Month	DD, YYYY}
	\revised{Month	DD, YYYY}
  \accepted{Month	DD, YYYY}
  \journalname{Nanophotonics}
  \journalyear{YYYY}
  \journalvolume{XX}
  \journalissue{X}
  \startpage{1}
  \aop
  \DOI{10.1515/sample-YYYY-XXXX}

\title{Parabolic-accelerating vector waves}

\author[1]{Bo Zhao}
\author[3]{Valeria Rodr\'iguez-Fajardo}
\author[1]{Xiao-Bo Hu}
\author[2]{Raul I. Hernandez-Aranda}
\author*[2]{Benjamin Perez-Garcia}
\author[1,4]{Carmelo Rosales-Guzm\'an$^\dagger$}
\runningauthor{Bo Zhao et al.}
\affil[1]{Wang Da-Heng Collaborative Innovation Center, Heilongjiang Provincial Key Laboratory of Quantum Manipulation and Control, Harbin University of Science and Technology, Harbin 150080, China}
\affil[2]{Photonics and Mathematical Optics Group, Tecnologico de Monterrey, Monterrey 64849, Mexico, e-mail: b.pegar@tec.mx}
\affil[3]{School of Physics, University of the Witwatersrand, Private Bag 3, Johannesburg 2050, South Africa}
\affil[4]{Centro de Investigaciones en Óptica, A.C., Loma del Bosque 115, Colonia Lomas del campestre, 37150 León, Gto., Mexico, e-mail: carmelorosalesg@cio.mx}
	
	
\abstract{Complex vector light fields have become a topic of late due to their exotic features, such as their non--homogeneous transverse polarisation distributions and the non-separable coupling between their spatial and polarisation degrees of freedom. In general, vector beams propagate in free space along straight lines, being the Airy-vector vortex beams the only known exception. Here, we introduce a new family of vector beams that exhibit novel properties that have not been observed before, such as their ability to freely accelerate along parabolic trajectories. In addition, their transverse polarisation distribution only contains polarisation states oriented at exactly the same angle but of different ellipticity. We anticipate that these novel vector beams might not only find applications in fields such as optical manipulation, microscopy or laser material processing, but extend to others.}


\maketitle

\section{Introduction} 
The ingrained notion that light travels along a straight line was first defied in 2007 by Siviloglou {\it et al.} \cite{Siviloglou2007a}, who introduced a novel kind of light beam with the ability to self-accelerate along a parabolic trajectory upon free space propagation \cite{Siviloglou2007b}. Such light beams, known as Airy beams, are natural solutions of the normalised paraxial wave equation. Crucially, even though they seem to propagate along parabolic trajectories, their centroid propagates along straight lines, in accordance to the electromagnetic momentum conservation law. Along with the discovery of Airy beams, their fascinating properties prompted the development of novel applications, which have impacted a wide diversity of fields, such as, optical manipulation, microscopy, laser material processing, among others (see for example \cite{Efremidis2019} for an extensive review). More importantly, the experimental demonstration of Airy beams ignited the quest for other kinds of accelerating beams \cite{Bandres2008Accelerating, Bandres2009, Greenfield2011, Zhang2012, RosalesGuzman2013Airy, Ruelas2014, Patsyk2018, Aleahmad2012}. Of particular interest is the case of accelerating parabolic beams, which form a complete and infinite orthogonal family of solutions of the normalised paraxial wave equation \cite{Bandres2008Accelerating, Davis2009}. Such beams also propagate in free space in a non diffracting way describing parabolic trajectories.     

Noteworthy, most of the work carried out with accelerating beams has only considered the case of homogeneously polarised beams, while the manipulation of other degrees of freedom is gaining popularity, giving rise to a more general class of beams generally known as structured light fields. This is the case of complex vector light beams, classically-entangled in their spatial and polarisation degrees of freedom, which feature a non-homogeneous polarisation distribution across the transverse plane\cite{Forbes2021,Roadmap, Rosales2018Review}. Such beams have gained popularity in recent time not only due to their unique traits, such as their quantum-like non separability \cite{konrad2019quantum, Eberly2016, toninelli2019concepts, forbes2019classically, Toppel2014}, but also due to the many applications they are pioneering \cite{Hu2019, BergJohansen2015, Ndagano2018, Ndagano2017, Otte2020, Sit2017}. In vector beams, their spatial and polarisation degrees of freedom are coupled in a non-separable way, which generates the non-homogeneous polarisation distribution. Importantly, while the polarisation degree of freedom is restricted to a two-dimensional space, the spatial one is not, as any of the unbounded solution sets of the wave equation, either in its exact or paraxial version, can be used. Examples of vector beams that have been experimentally demonstrated are Bessel, Laguerre-, Ince- and Mathieu-Gauss beams, amongst others, all of which propagate along straight trajectories \cite{Zhan2009,Dudley2013,Otte2018a,Yao-Li2020,Rosales2021}. Along this line, previous works have demonstrated the acceleration of vectorial fields, in which case, their polarisation structures rotates around the optical axis, while still propagate along straight lines \cite{Agela2021}. Perhaps the only known case of a vector beam capable to propagate along a parabolic trajectory is the Airy-vortex vector beam \cite{Zhou2015}.

We propose and experimentally demonstrate a new family of vector beams, which we term Accelerating Vector Waves (AVWs), that are non-separable weighted superpositions of the polarisation and spatial degrees of freedom encoded in the orthogonal set of accelerating waves. These beams exhibit two interesting properties, namely, that their non-homogeneous polarisation distributions propagate in free space along parabolic trajectories maintaining a maximum degree of coupling, and that, even though the non-homogeneous transverse polarisation distribution of an individual AVW contains different states of elliptical polarisation, all of them are located on a great circle on the Poincar\'e sphere representation for polarisation. Here, we start by describing these beams theoretically, then move to their implementation in the laboratory, and finally show experimental results to showcase their novel features. Due to their intriguing properties, we expect AVWs will attract the wide interest of the optical community, stemming not only from their potential applications but also from their fundamental aspects. 

\section{Theory}
\begin{figure}[t]
    \centering
    \includegraphics[width=.46\textwidth]{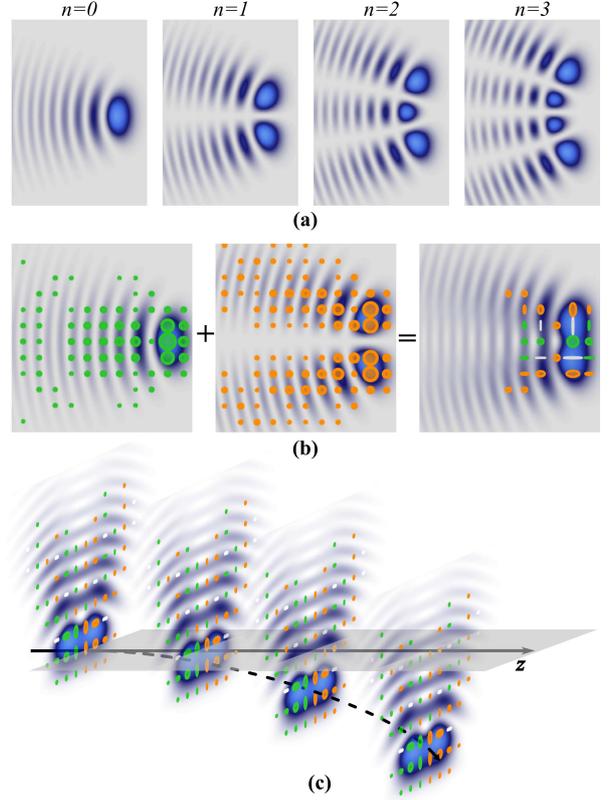}
    \caption{(a) Intensity profiles of accelerating parabolic scalar waves of orders $n=\{0, 1, 2, 3\}$. (b) Schematic representation of the non-separable superposition of two orthogonal scalar modes carrying orthogonal polarisations to generate an accelerating vector wave $\AVW{}_{n,m}(\eta,\xi,z)$. (c) Illustration of the free-space propagation trajectory of an AVW along the $z$ axis. Green and orange ellipses represent right- and left-handed circular polarisation, respectively, and white lines represent linear polarisation.}\label{fig:Concept}
\end{figure}

Accelerating parabolic waves (APWs) are solutions of the paraxial wave equation in parabolic coordinates. They are non-diffracting beams that accelerate during free-space propagation.  Their experimentally realisable finite--energy form is given by \cite{Bandres2008Accelerating}
\begin{align}
    \phi_n(\eta, \xi, z) =& \exp[i(z/2k\kappa^2 - ia)(\eta^2-\xi^2)/2]\times\\ \nonumber
    &\exp[i(z/2k\kappa^2-ia)^3/3]\Theta_n(\eta)\Theta_n(i\xi),
\end{align}
where the parabolic coordinates $(\eta, ~\xi)$ are related to the Cartesian coordinates by $(\eta^2/2 - \xi^2/2,~\eta\xi) = (x/\kappa-(z/2\kappa^2)^2+iaz/k\kappa^2, ~y/\kappa)$, $\kappa$ is a transverse scale parameter, $k$ is the wave number, $z$ is the propagation distance and $a$ is parameter that controls the exponential aperture of the beam at $z=0$. The functions $\Theta_n(\cdot)$ correspond to the solutions of the differential equation
\begin{align}\label{eq:quartic}
    \left(-\frac{1}{2}\dd{}{\eta}{2} + \frac{\eta^4}{4}\right)N(\eta) = E\:N(\eta),
\end{align}
which corresponds to the one--dimensional Schr\"odinger equation with potential $V(\eta) = \eta^4/4$ (known as quartic potential) and $m=\hbar=1$ \cite{Banerjee1978}. Importantly, the eigen-solutions $\Theta_n$ ($n\in \mathbb{N}$) of Eq. \ref{eq:quartic} form an orthogonal set of functions, whose parity is governed by $n$. Since these eigen-solutions cannot be expressed in a closed form, a suitable numerical method  must be employed to obtain them \cite{Driscoll2014}. In particular, we are interested in square integrable eigen-solutions of Eq.~\ref{eq:quartic}.  Fig.~\ref{fig:Concept}(a) shows the intensity profiles of the scalar APWs for $n=\{0,1,2,3\}$. As can be seen, for $n=0$ the intensity profile of the beam contains only one main lobe of maximum intensity and additional subsequent lobes of decaying intensity. In general, for $n>0$ the intensity profile is formed by $n+1$ lobes of maximum intensity.

Mathematically, the accelerating Vector Waves (AVWs) are constructed as a superposition of two scalar APWs with orthogonal polarisations, such that at the $z$--plane and for indices $n,m$ they are given by
\begin{align}\nonumber
    \AVW{}_{n,m}(\eta,\xi,z) =& \cos\alpha \:\phi_n(\eta,\xi,z)\hatt{e}_1(\theta,\varphi) + \\ \label{eq:avw}
    &\sin\alpha \exp(i\beta)\:\phi_m(\eta,\xi,z)\hatt{e}_2(\theta,\varphi),
\end{align}
where the weighting factor $\alpha\in[0,\pi/2]$ allows the field to change from scalar to vector, and the parameter $\beta=[0,\pi]$ controls the inter-modal phase. The basis vectors 
\begin{align}
    \hatt{e}_1(\theta,\varphi) = \cos(\theta/2)\hatt{e}_R + \exp(i\varphi)\sin(\theta/2)\hatt{e}_L,\\
    \hatt{e}_2(\theta,\varphi) = \sin(\theta/2)\hatt{e}_R - \exp(i\varphi)\cos(\theta/2)\hatt{e}_L,
\end{align}
represent the general elliptical polarisation basis. Note that we can obtain the left/right-handed circular polarisation basis by setting $\theta=\pi$ and $\varphi=0$ and the horizontal/vertical basis with $\theta=\pi/2$ and $\varphi=0$. Without loss of generality, here we will restrict our results to the circular polarisation basis, only briefly mentioning some theoretical examples of the horizontal/vertical basis. Fig.~\ref{fig:Concept}(b) illustrates conceptually the above description for the specific case $\AVW{}_{0,1}(\eta,\xi,z)$ with $\alpha=\pi/4$ and $\beta=0$ as polarisation distributions overlay onto their corresponding intensity profiles. Left and middle panels show the two scalar modes $\theta \:\phi_n(\eta,\xi,z)\hatt{e}_L$ and $\phi_m(\eta,\xi,z)\hatt{e}_R$ with Right Circular polarisation (RCP) and Left Circular Polarisation (LCP), respectively, represented by green and orange ellipses for the first and second case respectively. Notice the intensity patters of the scalar modes are different, as required to obtain vector modes. In a similar way, the right panel presents the non-separable superposition of both scalar modes (Eq.~\ref{eq:avw}). The parabolic trajectory described by AVWs can be seen schematically in Fig.~\ref{fig:Concept}(c) for $\AVW{}_{2,3}(\eta,\xi,z)$ propagating along the $z$ axis. Mathematically, this is expressed in the transverse shift $y_s = [z/(2k)]^2/\kappa^3$ \cite{Davis2008}, which is independent of the indices $n$, $m$.



\section{Experimental details}

\begin{figure}[t]
    \centering
    \includegraphics[width=.46\textwidth]{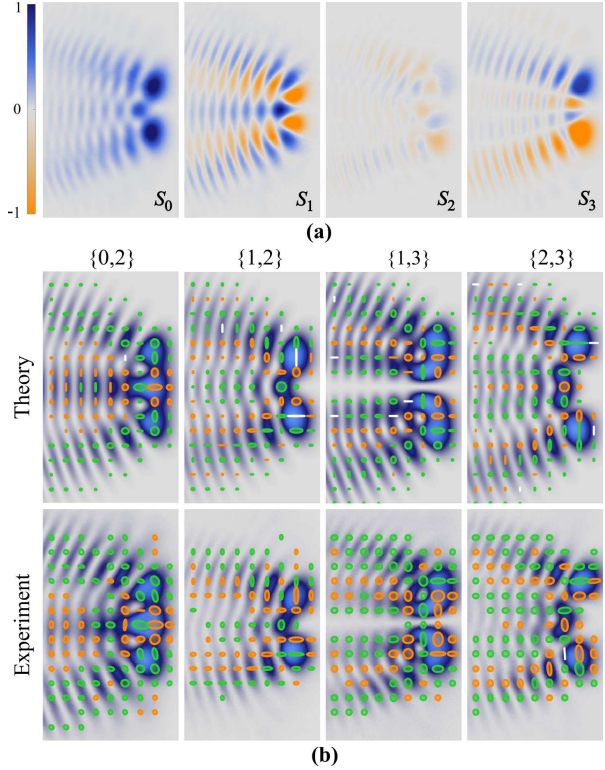}
    \caption{(a) Example of experimentally measured Stokes parameters $S_0$, $S_1$, $S_2$ and $S_3$ for $\AVW{}_{1,2}(\eta,\xi,z=0)$. (b) Theoretical and experimental reconstructed intensity and polarisation distribution from the Stokes parameters for the modes $\AVW{}_{0,2}(\eta,\xi,0)$, $\AVW{}_{1,2}(\eta,\xi,0)$, $\AVW{}_{1,3}(\eta,\xi,0)$ and $\AVW{}_{2,3}(\eta,\xi,0)$.}
    \label{fig:vectormodes}
\end{figure}

We implemented the AVW described above using a Digital Micromirror Device (DMD) and following the technique that we proposed and fully characterised in a previous article \cite{Rosales2020}. This device is polarisation-insensitive, very flexible and versatile, allowing the generation of vector modes with arbitrary spatial distributions, such as elliptical or parabolic \cite{Rosales2020,Rosales2021,Yao-Li2020,Xiaobohu2021}. In essence, a DMD is illuminated with two modes carrying orthogonal polarisations, impinging at slightly different angles but exactly at the geometric centre of the hologram displayed on the DMD. The hologram contains a superposition of the two transmittance functions that generate the constituting scalar modes of Eq.~\ref{eq:avw}, each with an additional unique linear spatial grating that redirects the mode along a specific angle, and whose periods are carefully chosen to guarantee both generated beams co-propagate along the same axis, where the desired vector beam is created. Transmittance functions are calculated as the inverse Fourier transform of the desired modes $\phi_n(\eta, \xi, z)$ \cite{Davis2008, Davis2009}, thus we add a lens in a $2f$ configuration, where $f$ is the focal length of the lens ($f=200$ mm in our case), and measure at its back focal plane. Intensity patterns of the generated beams were captured with a high resolution CCD camera (FL3-U3-120S3C-C with resolution of $4000\times3000$ pixels and a pixel size of 1.55 $\mu$m). Polarisation reconstruction was achieved through Stokes polarimetry, using a set of intensity measurements as detailed in \cite{Rosales2021}. Fig.~\ref{fig:vectormodes}(a) show\RH{s} an example of the experimentally measured Stokes parameters $S_0$, $S_1$, $S_2$ and $S_3$ for the specific mode  $\AVW{}_{1,2}(\eta,\xi,z=0)$. Reconstructed intensity and polarisation distributions of a set of representative examples of the experimentally generated $\AVW{}_{n,m}(\eta,\xi,z=0)$ modes using the circular polarisation basis are presented in Fig.~\ref{fig:vectormodes}(b), both for the theory (top row) and the experiment (bottom row). Notice the high similarity between the latter, demonstrating the good performance of our generation method.

\section{Results and discussion}
\begin{figure}[t]
    \centering
    \includegraphics[width=.48\textwidth]{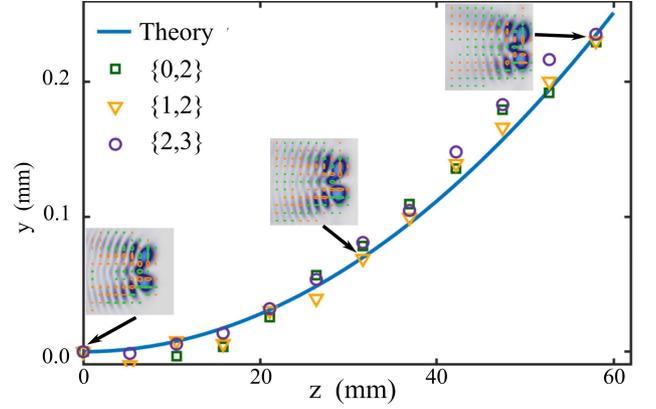}
    \caption{Shift of the $y$-coordinate of the maximum intensity lobe position for three AVWs upon free-space propagation. The continuous curve represents the expected position predicted by theory, whereas the data points correspond to experiment for the cases $\AVW{}_{0,2}(\eta,\xi,z)$ (squares), $\AVW{}_{1,2}(\eta,\xi,z)$ (triangles) and $\AVW{}_{2,3}(\eta,\xi,z)$ (circles). Notice that all three cases accelerate in an identical way. Insets show the transverse polarisation distribution overlapped with the intensity profile of the $\AVW{}_{1,2}(\eta,\xi,z)$ at three different planes.}
    \label{fig:trajectory}
\end{figure}

The vector modes described by equation \ref{eq:avw} and shown in Fig.~\ref{fig:vectormodes} propagate along parabolic trajectories maintaining not only their intensity and polarisation distribution but also a maximum coupling between both. We corroborated this by tracking the transverse spatial coordinates $(x,y)$ of one of the lobes of maximum intensity as function of their propagation distance $z$. We observed that while the $x$ coordinate remains almost constant, the $y$ coordinate shifts following a quadratic trend. Figure \ref{fig:trajectory}, in which the coordinate $y$ is plotted against the propagation distance $z$, clearly shows such behaviour for a representative set of AVWs given by $\AVW{}_{0,2}(\eta,\xi,z)$, $\AVW{}_{1,2}(\eta,\xi,z)$ and $\AVW{}_{2,3}(\eta,\xi,z)$. Since all modes shown where generated with the same initial parameters $k$ and $\kappa$, they accelerated in an identical way. Insets show examples of the polarisation distribution overlapped with the intensity distribution at three propagation distances $z=0$ mm, $z=20$ mm and $z=60$ mm for the mode $\AVW{}_{1,2}(\eta,\xi,z)$.

\begin{figure}[t]
    \includegraphics[width=.46\textwidth]{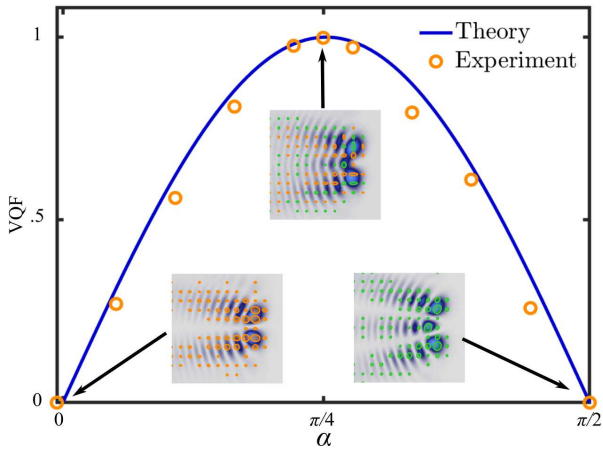}
    \caption{Vector Quality Factor (VQF) as a function of the weighing coefficient $\alpha\in[0,\pi/2]$ for the mode $\AVW{}_{2,3}(\eta,\xi,z=0)$. Insets show the intensity and polarisation distributions for $\alpha=0,\pi/4$, and $\pi/2$.}
    \label{fig:VQF}
\end{figure}

As mentioned earlier, AVWs can be generated with arbitrary degrees of non-separability, evolving from scalar to vector, via the parameter $\alpha$ (see Eq.~\ref{eq:avw}). More precisely, as $\alpha$ increases from 0 to $\pi/4$ the mode changes monotonically from a pure scalar mode with right-handed circular polarisation ($\alpha=0$) to a pure scalar mode with left-handed circular polarisation ($\alpha=\pi/2$), passing through a pure vector mode ($\alpha=\pi/4$). Intermediate values of $\alpha$ produce vector modes with intermediate degrees of non-separability, which can be measured through the concurrence or Vector Quality Factor (VQF), which is a measure borrowed from quantum mechanics that allows to quantify the degree of coupling between the spatial and polarisation degrees of freedom \cite{McLaren2015,Ndagano2015,Zhaobo2019}. Experimentally, the VQF can be quantified directly from the Stokes parameters as \cite{Selyem2019,Manthalkar2020},
\begin{equation}
VQF=\sqrt{1-\left(\frac{\mathbb{S}_1}{\mathbb{S}_0} \right)^2-\left(\frac{\mathbb{S}_2}{\mathbb{S}_0} \right)^2-\left(\frac{\mathbb{S}_3}{\mathbb{S}_0} \right)^2},
\label{concurrence}
\end{equation}
where $\mathbb{S}_i$ ($i=0,1,2,3$) is a number that results from integrating the Stokes parameters $S_i$ over the entire transverse profile, {\textit i. e.}, $\mathbb{S}_i=\iint_{-\infty}^\infty S_{i} dA$. Figure \ref{fig:VQF} shows a representative example of the VQF as function of $\alpha$ for the specific case $\AVW{}_{2,3}(\eta,\xi,z=0)$. As expected, the VQF increases from 0 to 1, as $\alpha$ increases from 0 to $\pi/4$ and then it decreases back to zero, as $\alpha$ reaches the value $\pi/2$. Insets show the intensity profile overlapped with the polarisation distribution for three key values, namely $\alpha=0, \pi/4$ and $\pi/2$.

\begin{figure}[h!]
    \centering
    \includegraphics[width=.46\textwidth]{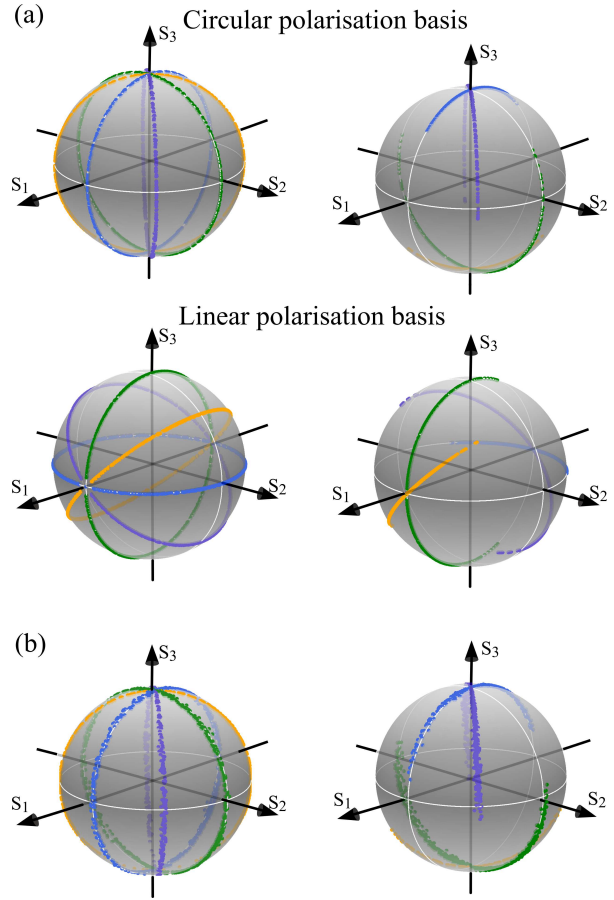}
    \caption{Geometric representation on the Poincar\'e sphere of the transverse polarisation distribution of the vector mode $\AVW{}_{0,2}(\eta,\xi,z=0)$. Left column shows modes with increasing values of the inter-modal phase, namely $\beta=0$ (blue), $\beta=pi/4$ (purple), $\beta=pi/4$ (green) and $\beta=3pi/4$ (yellow). Right column shows modes with different weighting coefficients, namely $\alpha=\pi/12, \pi/6, \pi/3$ and $4\pi/12$, with the same inter-modal phases as in the left column for the sake of better visibility. (a) Theoretical results for the case of AVW in the circular (top row) and linear (bottom row) polarisation basis. (b) Corresponding experimental results for the case of the circular basis.}
    \label{fig:Sphere}
\end{figure}

Finally, we analyse the polarisation states distribution of AVWs on the Poincar\'e sphere. It turns out that for a particular AVW all of them are mapped onto a great circle. Fig.~\ref{fig:Sphere}(a) shows numerical simulations in the circular (top row) and horizontal/vertical linear (bottom row) bases. It can be seen that in the first case, the great circles intersect the North and South poles, meaning all polarisation states in the AVW are oriented at exactly the same polarisation angle and only differ in their ellipticity, containing all polarisation states from circular to linear. Interestingly, a change in the inter-modal phase $\beta$ originates a rotation of the great circle around the $S_3$ axis (Fig.~\ref{fig:Sphere}(a) top left panel), and a change in the weighting coefficient $\alpha$ causes the polarisation states location from complete to incomplete great circles, in such a way that the arc length is proportional to it (Fig.~\ref{fig:Sphere}(a) top right panel).
Similarly, in the case of horizontal/vertical linear polarisation basis (Fig.~\ref{fig:Sphere}(a) bottom row), the great circles intersect the cross points between the $S_1$ axis and the sphere, and rotate around the $S_1$ axis when changing $\beta$. The effect of $\alpha$ is as in the circular basis.  For instance, for $\alpha=\pi/2$ and $\beta=0$, the polarisation distribution is mapped to the equator of the Poincar\'e sphere. For the case of circular polarisation basis, we corroborated this experimentally, as shown in Fig.~\ref{fig:Sphere}(b), that shows a remarkable similarity to its theoretical counterpart.
In summary, we have introduced theoretically and demonstrated experimentally a new family of vector beams with the ability to accelerate along parabolic trajectories upon free space propagation. Such accelerating beams differ quite dramatically from common vector beams, which always propagate along straight trajectories. These families of vector beams are constructed as a weighted superposition of the spatial and polarisation degrees of freedom carrying an inter-modal phase. To generate them, the spatial degree of freedom is encoded in a set of orthogonal solutions of the one--dimensional Schr{\"o}dinger equation with a quartic potential, known as accelerating waves. An important feature of such modes is their propagation-invariant spatial and polarisation structures, as we corroborated experimentally. Further, the weighting coefficient allows tuning from purely scalar to completely vectorial, passing through intermediate states, which was also corroborated experimentally using the well-known measure of concurrence from quantum mechanics adapted for vector beams. Another important feature of these accelerating vector modes lies in their transverse polarisation distribution, which is mapped onto great circles on the Poincar\'e sphere. In particular, in the circular polarisation basis the great circles intersect the North and South poles, and contain states of polarisation from linear to circular, all with the same ellipticity angle. For comparison, cylindrical vector modes are mapped either to an equatorial line or to the whole Poincar\'e sphere (known as full-Poincar\'e modes). Noteworthy, the inter-modal phase allows to rotate the circle of polarisation around the Poincar\'e sphere, leaving the points on North and South poles fixed. Given their interesting properties, we expect AVWs to find applications in fields such as optical manipulation, laser material processing, among others.


\begin{funding}
  Consejo Nacional de Ciencia y Tecnología (PN2016-3140); National Natural Science Foundation of China (61975047).
\end{funding}


\end{document}